# Paramagnetic Meissner effect in single crystal $Yb_3Rh_4Sn_{13}$

P. Kulkarni

*Department of Condensed Matter Physics and Materials Science, Tata Institute of Fundamental Research Colaba, Mumbai 400005*
*prasanna1609@gmail.com*

**Abstract.** Paramagnetic Meissner effect was observed in single crystal $Yb_3Rh_4Sn_{13}$. While field cooling the sample, at the onset of the superconducting transition at 7.92 K, the DC magnetization first goes through a minimum at 7.85 K, followed by the peak of the PME signal at 7.6 K and then crosses over to the diamagnetic state. The magnetization vs field curves are reversible above the minimum and imbibe a pinning characteristic below 7.85 K. The minimum can be attributed to the opposing contribution to the dc magnetization signal from surface superconductivity. The subsequent flux compression while cooling the sample below 7.85 K shows the rapid increase in the dc magnetization signal. PME signal is actually the diamagnetism opposing the flux compression.

**Keywords:** Paramagnetic Meissner effect, surface superconductivity
**PACS:** 74.25.Ld, 74.25.Ha

The positive dc magnetization signal below the superconducting transition was termed as paramagnetic Meissner effect (PME) [1-9]. It is counterintuitive as the superconductors contribute diamagnetic signal in the dc magnetization measurements. PME is significant at low fields, and suppresses with the increase in the applied magnetic field. The observation of PME in s-wave superconductors, with low pinning characteristics, was attributed to the flux trapping and compression. The role of surface superconductivity was also considered important for the observation of PME [9-15].

Recently, in the cubic stannide, $Ca_3Rh_4Sn_{13}$ ($T_c$ = 8.35 K), the detailed dc and ac magnetization studies of PME were reported [14]. For this compound, the magnetization response at very low fields was fluctuating and it could be related the presence of giant vortex state and its subsequent transformation to the Abrikosov vortex lattice predicted in the literature [5,12,13]. However, in these measurements the sign of the PME signal remained intriguing. If the external magnetic field is notionally negative, the diamagnetic superconducting state in the magnetization measurements gives positive signal. However, the PME signal does not change sign, and retains the same sign as the diamagnetic signal in the negative fields. In this case, the PME signal appears as superposing to the diamagnetic signal.

The quantum state of the trapped flux was studied using the Ginzburg-Landau equations [12,13] and it was shown that the instead of the Meissner state, the giant vortex state with orbital quantum number L > 0 would have the lower energy at the superconducting transition. As the sample is cooled the giant vortex would successively transformed into different lower energy states, and the magnetization signal would fluctuate before settling in the diamagnetic state [14].

The positive field cooled magnetization (PMFC) in high purity single crystal Nb sphere was reported earlier by Das et al. [9]. The results suggested that the surface superconductivity co-exist with the PMFC. The results in $Ca_3Rh_4Sn_{13}$ motivated further studies of conventional superconductors at low fields, and the detailed investigation of the normal to superconductor transition in $Yb_3Rh_4Sn_{13}$ reported in this paper.

## EXPERIMENTAL

A single crystal sample of $Yb_3Rh_4Sn_{13}$ prepared by the tin flux method was used for low field dc magnetization measurements. The crystal was mounted in the straw which contributes negligible magnetization signal. The dc magnetization measurements were performed using a commercial SQUID-Vibrating Sample Magnetometer (Quantum Design (QD) Inc., USA, model S-VSM). In S-VSM, the sample executes a small vibration around a mean position, where the magnetic field is uniform and maximum. This avoids the possibility of the sample moving in an inhomogeneous field during the dc magnetization measurements. The remnant field of the superconducting magnet of S-VSM was carefully estimated at different stages of the experiment, using a standard paramagnetic Palladium specimen.

## RESULTS AND DISCUSSIONS

Figure 1 shows the temperature dependence of the magnetization in single crystal $Yb_3Rh_4Sn_{13}$ at H = 50 Oe. The magnetization is measured during the cooling mode while the field is applied at T = 20 K, well above the superconducting transition $T_c$ = 7.92 K in $Yb_3Rh_4Sn_{13}$. The inset in figure 1 shows the field cooled cool-down (FCC) magnetization in the temperature range from 7.8 K to 7.91 K. At the onset of the superconducting transition, the magnetization decreases till the temperature reaches around 7.85 K, where the fall of magnetization slows down, and reverses. The magnetization is seen to rise rapidly below 7.84 K, and continues to peak at 7.66 K. If cooled further, the magnetization crosses over to zero value and reaches negative diamagnetic state below 5 K. The FCC magnetization curve can be considered as consisting of three regions, from $T_c$ down to 7.85 K, initial fall, paramagnetic Meissner signal from 7.84 K to 7.2 K and usual diamagnetic signal down to 3 K.

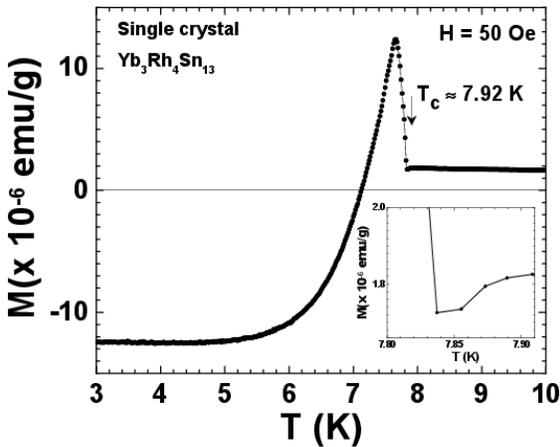

**FIGURE 1.** Main panel shows the magnetization vs temperature in the field cooled cooling at H = 50 Oe in single crystal $Yb_3Rh_4Sn_{13}$. The field is applied at 20 K and the superconducting transition temperature is marked at 7.92 K.

We measured the magnetization vs magnetic field at several temperatures in the temperature range from 7.7 to 8 K. In each case the sample was cooled initially from 10 K down to the selected temperature at applied magnetic field of 500 Oe. The M vs H curve is traced within H = +500 to -500 Oe and the sample was subsequently warmed upto 10 K. The M vs H at a temperature interval from 7.94 to 7.8 K were measured at every 0.1 K step, after field cooling in 500 Oe every time from T = 10 K. The M vs H curve at T = 7.94 K is linear, and the offset of the field is less than 2 Oe. The derivative of the M vs H curves at lower temperatures are shown in Fig. 2(a), in order to enable the monitoring of the onset of the superconducting transition. At T = 7.92 K, a small signal pertaining to the superconducting transition was observed, as can be seen in the derivative curve in Fig. 2(a). As the temperature is lowered the magnetization signal gains strength, and the derivative curves in Fig. 2(b) to 2(h), shows that the dc magnetization is reversible in the temperature range from 7.91 K to 7.85 K. This coincides with the temperature at which the minimum in the M vs T curve was observed in the inset panel in Figure 1.

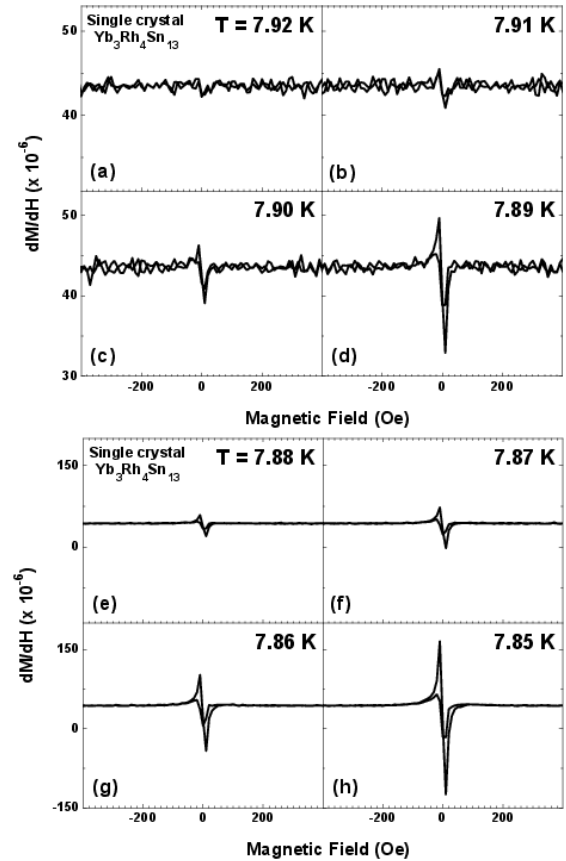

**FIGURE 2.** Panels (a) to (h) in the upper and lower set of panels show the derivative of the magnetization vs magnetic field curves between H = +500 to -500 Oe. In each panel, the temperature is marked which starts from T = 7.92 K to 7.85 K. The sample is field cooled in +500 Oe in each case.

In Figure 3(a-d), the M vs H are shown at temperature from 7.84 K to 7.81 K. The magnetization loops at 7.84 K in panel (a), start to open up, while the field is swept from -500 Oe to +500 Oe. At lower temperatures, the second branch of the loop develops further, as can been seen at 7.81 K in panel (d).

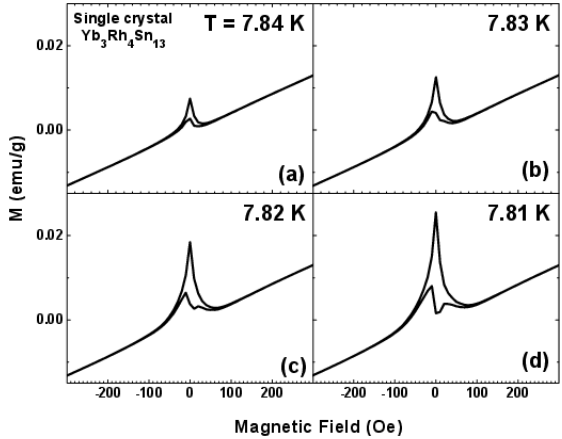

**FIGURE 3.** Panels (a) to (d) show the magnetization vs magnetic field at T = 7.84 K to 7.81 K. The sample is cooled in each case at +500 Oe down to the selected temperature and the loop is traced between +500 Oe to -500 Oe.

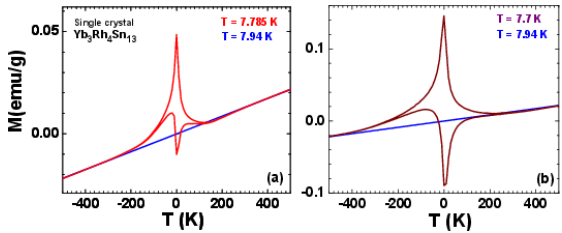

**FIGURE 4.** Panels (a) and (b) show the magnetization vs magnetic field at T = 7.785 and 7.7 K, respectively. In each panel the M vs H at 7.94 K is plotted for comparison.

In Figure 4 (a), we show the M vs H at 7.785 K, which is on the rising part of the M vs T curve in Figure 1. The linear M vs H at 7.94 K, is also plotted in panel (a), for a comparison. At 7.7 K, the M vs H is shown in panel (b) along with the linear curve at 7.94 K. In both these panels, M vs H is considerably open, and at 7.7 K in particular, the typical magnetization behavior with the magnetic field for a superconductor with low pinning is seen.

The new observations reported in this paper in single crystal $Yb_3Rh_4Sn_{13}$ suggest that at the onset of the superconducting transition the dc magnetization signal in the normal state is opposed by the surface superconductivity. As the superconductivity percolates inside the bulk and consequently compressing the flux, another opposing dc magnetization signal is produced, which has the same sign as the external field. This produces a peak in the dc magnetization signal leading to the paramagnetic Meissner effect. These results show previously unknown co-existence of paramagnetism and superconductivity near the transition. The flux expulsion may be associated with the diamagnetic signal in the dc magnetization measurements, while the flux compression results in the positive signal.

## ACKNOWLEDGMENTS

Author acknowledges the SQUID-VSM facility at TIFR provided by A. K. Grover.